\documentclass[pra,twocolumn,showpacs,twoside]{revtex4}

\usepackage{amssymb, amsbsy, amsmath, latexsym, dsfont, array, layout, graphicx}
\usepackage{color}

\newcommand{\ket}[1]{\left|{#1}\right\rangle}
\newcommand{\bra}[1]{\left\langle{#1}\right|}

\begin{document}
\title{Universal quantum computing with semiconductor double-dot molecules on a chip}
\author{Peng Xue}
\affiliation{Department of Physics, Southeast University, Nanjing
211189, P. R. China}
\affiliation{Institute for Quantum Information
Science, University of Calgary, Alberta T2N 1N4, Canada}
\date{\today}

\begin{abstract}
We develop a scalable architecture for quantum computation using
controllable electrons of double-dot molecules coupled to a
microwave stripline resonator on a chip, which satisfies all
Divincenzo criteria. We analyze the performance and stability of all
required operations and emphasize that all techniques are feasible
with current experimental technologies.
\end{abstract}


\maketitle

\section{Introduction}
Quantum computing enables some computational
problems to be solved faster than would ever be possible with a
classical computer~\cite{Gro97} and exponentially speeds up
solutions to other problems over the best known classical
algorithms~\cite{Sho94}. Of the promising technologies for quantum
computing, solid-state implementations such as spin qubits in
quantum dots~\cite{LD97} and bulk silicon~\cite{Kan98}, and charge
qubits in bulk silicon~\cite{ABW+07} and in superconducting
Josephson junctions~\cite{WSB+05}, are especially attractive because
of stability and expected scalability of solid-state systems; of
these competing technologies, semiconductor double-dot molecules
(DDMs) are particularly important because of the combination spin
and charge manipulation to take advantage of long memory times
associated with spin states and at the same time to enable efficient
readout and coherent manipulation of charge states.

Here we develop a scalable architecture for semiconductor quantum
computation based on two-electron states in
DDMs~\cite{Im99,Petta05,Johnson05,GI06,Taylor07} coupled to a
microwave stripline resonator~\cite{Taylor06,Guo09}. The quantum
information is encoded in the superpositions of double-dot singlet
states. The initialization of qubit states can be implemented by an
adiabatic passage. A universal set of gates including single- and
two-qubit gates can be implemented via the resonator-assisted
interaction with a microwave stripline resonator and the requirement
for electrically driving DDMs directly is released, which avoids
moving the system away from the optimal point (where the coupling is
achieved to be strongest) because of the potential difference caused
by the electric drive and increasing the extended dephasing due to
the fluctuations of the electric field. Compared to the previous
protocol~\cite{Guo09}, the system in our scheme always works in the
strong coupling regime and the second order dephasing time is
considered here. The readout of qubits can be realized via microwave
irradiation of the resonator by probing the transmitted or reflected
photons. The main decoherence processes are dissipation of the
stripline resonator, charge-based relaxation and dephasing of the
semiconductor DDMs, and spin dephasing limited by hyperfine
interactions with nuclei. By numerical analysis we show all gate
operations and measurements can be implemented within the coherent
life time of qubits. Thus we address all Divincenzo
criteria~\cite{DiVincenzo} and show all play important roles in the
dynamics of the two-electron system but none represents a
fundamental limit for quantum computing.

\section{Qubits}

\begin{figure}[tbp]
   \includegraphics[width=7.5cm]{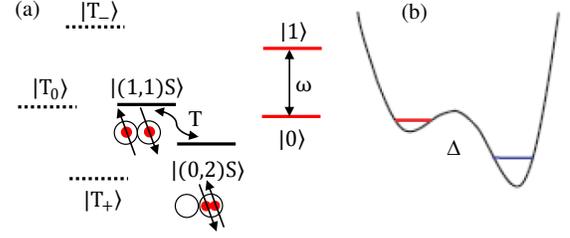}
   \caption{(a) Energy level diagram showing the $(0,2)$ and $(1,1)$ singlets
   in solid lines and the three $(1,1)$ triplets in dashed lines, and the logical qubit
   states $\ket{0}$ and $\ket{1}$ in red lines. (b) Schematic of the double-well
   potential with an energy offset $\Delta$ provided by the external electric
   field along $x$ axis.}
   \label{fig:energylevel}
\end{figure}

We consider the system with two electrons located
in adjacent quantum dots coupling via tunneling. Imagine one of the
dots is capacitively coupled to a stripline
resonator~\cite{GI06,Taylor06,Guo09}. With an external magnetic
field $B_z=100$mT along $z$ axis, the spin aligned states
$\ket{T_+}=\ket{\uparrow\uparrow}$ and
$\ket{T_-}=\ket{\downarrow\downarrow}$, and the spin-anti-aligned
states
$\ket{T_0}=(\ket{\downarrow\uparrow}+\ket{\downarrow\uparrow})/\sqrt{2}$
and
$\ket{(1,1)S}=(\ket{\downarrow\uparrow}-\ket{\downarrow\uparrow})/\sqrt{2}$
have energy gaps due to the Zeeman splitting shown in
Fig.~\ref{fig:energylevel}(a). The notation $(n_L,n_R)$ labels the
number of electrons in the left and right quantum dots. The doubly
occupied state $\ket{(0,2)S}$ is coupled via tunneling $T$ to the
singlet state $\ket{(1,1)S}$. The double-dot system can be described
by an extended Hubbard Hamiltonian
$\hat{H}=(E_\text{os}+\mu)\sum_{i,\sigma}\hat{n}_{i,\sigma}-T\sum_{\sigma}(\hat{c}^\dagger_{L,\sigma}\hat{c}_{R,\sigma}+\text{hc})
+U\sum_i\hat{n}_{i,\uparrow}\hat{n}_{i,\downarrow}+W\sum_{\sigma,\sigma'}\hat{n}_{L,\sigma}\hat{n}_{R,\sigma'}+\Delta\sum_{\sigma}(\hat{n}_{L,\sigma}-\hat{n}_{R,\sigma})$
for $\hat{c}_{i,\sigma}$ ($\hat{c}^\dagger_{i,\sigma}$) annihilating
(creating) an electron in quantum dot~$i\in\{L,R\}$ with
spin~$\sigma\in\{\uparrow,\downarrow\}$,
$\hat{n}_{i,\sigma}=\hat{c}^\dagger_{i,\sigma}\hat{c}_{i,\sigma}$ a
number operator, and $\Delta$ an energy offset yielded by the
external electric field along $x$ axis shown in
Fig.~\ref{fig:energylevel}(b). The first term corresponds to on-site
energy~$E_\text{os}$ plus site-dependent field-induced
corrections~$\mu$. The second term accounts for $i \leftrightarrow
j$ electron tunneling with rate~$T$, and the third term is the
on-site charging cost~$U$ to put two electrons with opposite spin in
the same dot, and the fourth term corresponds to inter-site Coulomb
repulsion. In the basis $\{\ket{(1,1)S},\ket{(0,2)S}\}$, the
Hamiltonian can be deduced as
\begin{equation}
\hat{H}_\text{d}=-\Delta\ket{(0,2)S}\bra{(0,2)S}+T\ket{(1,1)S}\bra{(0,2)S}+\text{hc}.
\label{eq:deduceHam1}
\end{equation}

With the energy offset $\Delta$, degenerate perturbation theory in
the tunneling $T$ reveals an avoided crossing at this balanced point
between $\ket{(1,1)S}$ and $\ket{(0,2)S}$ with an energy gap
$\omega=\sqrt{\Delta^2+4T^2}$, and the effective tunneling between
the left and right dots with the biased energies $\Delta$ is changed
from $T$ to $T'=\omega/2$. We choose the superpositions of the
singlet states as our qubit states:
\begin{align}
&\ket{0}\equiv(\ket{(1,1)S}-\ket{(0,2)S})/\sqrt{2};\nonumber\\
&\ket{1}\equiv(\ket{(1,1)S}+\ket{(0,2)S})/\sqrt{2}.
\label{eq:qubit}
\end{align}

\begin{figure}[tbp]
   \includegraphics[width=8.5cm]{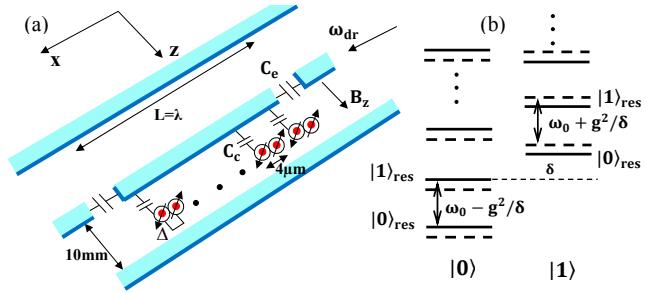}
   \caption{(a) Schematic of DDMs, biased with an energy offset $\Delta$,
   capacitively coupled to the stripline resonator. The coupling can be switched
   on and off via the external electric field along $x$ axis. The stripline
   resonator is driven by a classical field along $x$ axis.
   (b) Energy spectrum of the bared (solid lines) and dressed (dashed lines)
   states in the dispersive regime.}
   \label{fig:schematic}
\end{figure}

The essential idea is to use an effective electric dipole moment of
the singlet states $\ket{(1,1)S}$ and $\ket{(0,2)S}$ of a DDM
coupled to the oscillating voltage associated with a stripline
resonator shown in Fig.~\ref{fig:schematic}(a). We consider a
stripline resonator with length $L$, the capacitance coupling of the
resonator to the dot $C_\text{c}$, the capacitance per unit length
$C_0$, the total capacitance of the double-dot $C_\text{tot}$, and
characteristic impedance $Z_0$. The fundamental mode frequency of
the resonator is $\omega_0=\pi/LZ_0C_0$. The resonator is coupled to
a capacitor $C_e$ for writing and reading the signals. Neglecting
the higher modes of the resonator and working in the rotating frame
with the rotating wave approximation, we obtain an effective
interaction Hamiltonian as
\begin{equation}
\hat{H}_\text{int}=g(\hat{a}\hat{\sigma}_++\text{hc})
\label{eq:intHam}
\end{equation}
with $\hat{a}$ ($\hat{a}^\dagger$) the annihilation (creation)
operator of the resonator field, $\hat{\sigma}_+=\ket{1}\bra{0}$,
$\hat{\sigma}_-=\ket{0}\bra{1}$, and the effective coupling
coefficient
\begin{equation}
g=\frac{1}{2}e\frac{C_\text{c}}{LC_\text{tot}C_0}\sqrt{\frac{\pi}{Z_0}}\sin2\theta
\label{eq:g}
\end{equation}
with $\theta=\frac{1}{2}\tan^{-1}(\frac{2T}{\Delta})$.

The interaction between the resonator and qubit states is switchable
via tuning the electric field along $x$ axis. In the case of the
energy offset yielded by the electric field $\Delta\approx 0$, we
obtain the maximum value of the coupling between the resonator and
singlets in double dots. That is so called the optimal point.
Whereas $\Delta\gg T$, $\theta$ tends to $0$, the interaction is
switched off.

\section{Initialization and transportation}
Initialization of qubit
states can be implemented by an adiabatic passage between the two
singlet states~\cite{Taylor07}. Controllably changing $\Delta$
allows for adiabatic passage to past the charge transition, with
$\ket{(0,2)S}$ as the ground state if $\Delta \gg T$ achieved. First
we turn on the external electric field along $x$ axis and prepare
the two electrons of DDMs in the state $\ket{(0,2)S}$ by a large
energy offset $\Delta$. We change $\vartheta$ adiabatically to
$\pi/4$ by tuning the electric field, and then initialize the qubits
in the qubit state $\ket{0}$.

The SWAP operation~\cite{Taylor06}, where a qubit state is swapped
with a photonic state of the resonator, can be used to implement
transmission of qubits. If there is no photon in the resonator, with
the detuning $\delta=|\omega-\omega_0|=0$ and evolution time
$\pi/g$, a qubit is mapped to the photonic state in the resonator
$(\alpha\ket{0}+\beta\ket{1})\ket{0}_\text{res}\longrightarrow
\ket{0}(\alpha\ket{0}+\beta\ket{1})_\text{res}.$ Then we switch off
the coupling between this qubit and resonator and switch on that
between the desired qubit and resonator via the local electric
fields along $x$ axis. After the same evolution time, the previous
qubit state is transmitted to the desired qubit via the interaction
with the resonator.

\section{A universal set of gates}
Single-qubit gates including
bit-flip and phase gates, and an entangling two-qubit gate can be
implemented via resonator-associated interaction with a stripline
resonator. We consider a DDM interacts with a stripline resonator
field, which is driven by a strong classical field along $x$ axis
\begin{equation}
\hat{H}_\text{dr}(t)=\Omega(\hat{a}^\dagger e^{-\text{i}\omega_\text{dr}t}+\text{hc})
\label{eq:drHam}
\end{equation}
with the Rabi frequency $\Omega$, and the frequency of the classical
field $\omega_\text{dr}$ substantially detuned from the resonator
frequency $\omega_0$. In the rotating frame at the frequency
$\omega_\text{dr}$ for a single qubit and the field, we obtain
\begin{equation}
\hat{H}_{1q}=(\omega_0-\omega_\text{dr})\hat{a}^\dagger\hat{a}+\frac{\omega-\omega_\text{dr}}{2}\hat{\sigma}_z-g(\hat{a}^\dagger\sigma_-+\text{hc})+\frac{\Omega_\text{R}}{2}\hat{\sigma}_x
\label{eq:1qHam}
\end{equation}
with the effective Rabi frequency $\Omega_\text{R}=\frac{2\Omega g}{\omega_0-\omega_\text{dr}}$.

In the dispersive regime $\delta\gg g$, we obtain the effective
Hamiltonian from Eq.~(\ref{eq:1qHam}) as
\begin{equation}
\hat{H}_x=(\omega_0-\omega_\text{dr})\hat{a}^\dagger\hat{a}+\frac{\omega+g^2/\delta-\omega_\text{dr}}{2}\hat{\sigma}_z+\frac{\Omega_\text{R}}{2}\hat{\sigma}_x.
\label{eq:xHam}
\end{equation}
By choosing $\omega_\text{dr}=\omega+g^2/\delta$, the Hamiltonian
(\ref{eq:xHam}) evolves as a rotation around the $x$ axis. The gate
operates on the time scaling $t_x\sim1/\Omega_\text{R}$. These Rabi
oscillations have already been observed experimentally
in~\cite{WSB+05}.

In another case the drive is sufficiently detuned from the qubit
$|\omega-\omega_\text{dr}|\gg\Omega_\text{R}$~\cite{Blais07}, we
obtain a different effective Hamiltonian from Eq.~(\ref{eq:1qHam})
\begin{equation}
\hat{H}_z=(\omega_0-\omega_\text{dr})\hat{a}^\dagger\hat{a}
+\frac{\Omega'_\text{R}}{2}\hat{\sigma}_z,
\label{eq:zHam}
\end{equation}
which generates rotations around $z$ axis at a rate
$\Omega'_\text{R}=\omega+g^2/\delta-\omega_\text{dr}+\frac{1}{2}\frac{\Omega_\text{R}^2}{\omega-\omega_\text{dr}}$.
The time scaling of this gate operation is
$t_z\sim1/\Omega'_\text{R}$.

Since we can switch on and off the coupling between the resonator
and any DDM by tuning the local electric fields along $x$ axis, for
the case of two identical DDMs simultaneously coupled to the
resonator, without the drive, in the dispersive regime, we obtain
the effective Hamiltonian for the system from Eq.~(\ref{eq:intHam})
\begin{align}
\hat{H}_{2q}=&(\omega_0+\frac{g^2}{\delta}\sum_{i=1,2}\hat{\sigma}_z^i)\hat{a}^\dagger\hat{a}
+\frac{1}{2}(\omega+\frac{g^2}{\delta})\sum_{i=1,2}\hat{\sigma}_z^i\nonumber\\
&+\frac{g^2}{\delta}(\hat{\sigma}^1_+\hat{\sigma}_-^2+\text{hc}).
\label{eq:2qHam}
\end{align}
In the rotating frame at the frequency $\omega$, the evolution
operation of the two-qubit system dominated by the above Hamiltonian
after tracing out the field state (assume there is one photon in the
resonator)
\begin{equation}
U_{2q}=\exp{\Big\{-\text{i}\frac{g^2}{\delta}t\frac{3}{2}\sum_{i=1,2}\hat{\sigma}_z^i\Big\}}\sqrt{\text{iSWAP}}
\end{equation}
which provides an entangling two-qubit gate---root of SWAP gate on a
time scaling $t_{2q}$ that satisfies $g^2t_{2q}/\delta=\pi/4$.

Hence we have built a universal set of gates for quantum computing
with semiconductor DDMs coupled to a stripline resonator field.
Compared to the previous protocols~\cite{Guo09,Zheng02}, we drive
the resonator instead of driving qubits directly to implement
single-qubit gates, in which no addressing qubits individually is
required. The feasibility of single-qubit gates has already been
proved in~\cite{WSB+05} experimentally. For the two-qubit gate, we
realize it with the off-resonant interaction between both qubits and
resonator and release the requirement for driving qubits with a
strong classical field~\cite{Guo09,Zheng02}. Compared to the
previous protocol~\cite{Guo09}, the system works in the strong
coupling regime and the second order dephasing time is considered.
Driving qubits introduces large energy difference between the
potentials which moves the system away from the optimal point as
presented in~\cite{Guo09}. In that case the interaction is out of
the strong coupling regime even with the maximum coupling, e.g.
$gT_2<1$ with $T_2$ the extended dephasing time.

\section{Readout}
To perform a measurement of qubits, a drive of
frequency $\omega_\text{dr}$ modeled by Eq.~(\ref{eq:drHam}) is sent
through the resonator. In the dispersive regime, the energy gap
between the dressed states $\ket{0}_\text{res}$ and
$\ket{1}_\text{res}$ is $\omega_0-g^2/\delta$ for the qubit in the
state $\ket{0}$, while the energy gap $\omega_0+g^2/\delta$ for the
state $\ket{1}$ shown in Fig.~\ref{fig:schematic}(b).  The matrix
element of the Hamiltonian $\hat{H}_\text{dr}$ corresponding to a
bit-flip from the state $\ket{1}$ is suppressed, and depending on
the qubit being in the states $\ket{0}$ or $\ket{1}$ the
transmission spectrum will present a peak of width $\kappa$ (the
resonator decay rate) at $\omega_0-g^2/\delta$ or
$\omega_0+g^2/\delta$. This dispersive pull of the resonator
frequency is $\pm g^2/\kappa\delta$, and the pull is power dependent
and decreases in magnitude for photon numbers inside the
resonator~\cite{Blais04}. Via microwave irradiation of the resonator
by probing the transmitted or reflected photons, the readout of
qubits can be realized and completed on a time scaling
$t_m=1/\gamma_\phi$, where
$\gamma_\phi=8\bar{n}(\frac{g^2}{\delta})^2\frac{1}{\kappa}$ is the
dephasing rate due to quantum fluctuations of the number of photon
$\bar{n}$ within the resonator.

\section{Decoherence} Now we analyze the dominant noise source of our
system including the charge-based dephasing and relaxation, the spin
phase noise due to hyperfine coupling and the photon loss. Coupling
to a phonon bath causes relaxation of the charge system in a time
$T_1$. The characteristic charge dephasing with a rate $T_2^{-1}$.
The time-ensemble-averaged dephasing time $T_2^*$ is limited by
hyperfine interactions with nuclear spins. The decay of the
resonator $\kappa$ is considered as another dominant source of
decoherence.

For the charge relaxation time $T_1$, the decay is caused by
coupling qubits to a phonon bath. With the spin-boson model, the
perturbation theory gives an overall error rate from the relaxation
and incoherent excitation, with which one can estimate the
relaxation time $T_1\sim1\mu$s~\cite{Taylor06}.

The charge dephasing $T_2$ rises from variations of the energy
offset $\delta(t)=\delta+\epsilon(t)$ with
$\langle\epsilon(t)\epsilon(t')\rangle=\int d\omega
S(\omega)e^{\text{i}\omega(t-t')}$, which is caused by the low
frequency fluctuation of the electric field. The gate bias of the
qubit drifts randomly when an electron tunnels between the metallic
electrode. Due to the low frequency property, the effect of the
$1/f$ noise on the qubit is dephasing rather than relaxation. At the
zero derivative point, compared to a bare dephasing time
$T_b=1/\sqrt{\int d\omega S(\omega)}$, the charge dephasing is
$T_2\sim \omega T_b^2$ near the optimal point $\delta=0$. The bare
dephasing time $T_b\sim 1$ns was observed in~\cite{Hayashi04}. Then
the charge dephasing is estimated as $T_2\sim10-100$ns. Using
quantum control techniques, such as better high- and low-frequency
filtering of electronic noise, $T_b$ exceeding $1\mu$s was
observed~\cite{Petta05}, which suppresses the charge dephasing.

The hyperfine interactions with the gallium arsenide host nuclei
causes nuclear spin-related dephasing $T_2^*$. The hyperfine field
can be treated as a static quantity, because the evolution of the
random hyperfine field is several orders slower than the electron
spin dephasing. In the operating point, the most important
decoherence due to hyperfine field is the dephasing between the
singlet state $\ket{(1,1)S}$ and one of the triplet state
$\ket{T_0}$. By suppressing nuclear spin fluctuation, the dephasing
time can be obtained by quasi-static approximation as
$T_2^*=1/g\mu_B\langle\Delta B_n^z\rangle_\text{rms}$, where $\Delta
B_n^z$ is the nuclear hyperfine gradient field between two coupled
dots and rms means a root-mean-square time-ensemble average. A
measurement of the dephasing time $T_2^*\sim10$ns was demonstrated
in~\cite{Petta05}.

The quality factor $Q$ of the superconducting resonator in the
microwave domain can be achieved $10^6$~\cite{Wallraff04}. In
practice, the local external magnetic field $\sim 100$mT reduces the
limit of the quality factor to $Q\sim 10^4$~\cite{Frunzio05}. The
dissipation of the resonator $\kappa=\omega_0/Q$ leads the decay
time about $1\mu$s with the parameters $\omega_0=2\pi\times10$GHz.

\section{Feasibility} Now we analyze the feasibility of the proposal
with a gate-defined double-dot device as an example which is
fabricated using a GaAs/AlGaAs heterostructure grown by molecular
beam epitaxy with a two-dimensional electron gas $100$nm below the
surface, with density ~$2\times10^{11}$cm$^2$~\cite{Petta05}. When
biased with negative voltages, the patterned gates create a
double-well potential shown in Fig.~\ref{fig:energylevel}(b). The
quantum-mechanical tunneling $T$ between the two quantum dots is
about $T\simeq0-10\mu$eV. The stripline resonator can be fabricated
with existing lithography techniques~\cite{Wallraff04}. The qubit
can be placed within the resonator formed by the transmission line
to strongly suppress the spontaneous emission. The stripline
resonator in coplanar waveguides with $Q\sim10^4$ have already been
demonstrated in~\cite{Frunzio05}. The diameter of the quantum dot is
about $400$nm, and the corresponding capacitance of the double-dot
$C_\text{tot}$ is about $200$aF. The capacitive coupling of the
resonator to the dot is about
$C_\text{c}\approx2C_\text{tot}=400$aF. In practice, for
$\omega_0=2\pi\times10$GHz, $Z_0=50\Omega$, $L\sim \lambda=3$cm, the
coupling coefficient $g\sim 2\pi\times125$MHz is achievable by the
numerical estimations in~\ref{eq:g}. The frequency and coupling
coefficient can be tuned by changing $LC_0$. The external magnetic
field along $z$ axis is about $B_z=100$mT to make sure the energy
splitting $E_z=g\mu_B B_z$ between the two triplet states
$\ket{T_{\pm}}$ is larger than $\omega\sim\omega_0$.

With these parameters we can estimate the time scaling for quantum
computing. The time for transmitting a qubit to a photonic qubit in
the resonator is about $t_\text{tr}=\pi/g\approx4$ns. Readout of
qubits takes the time $t_m\approx0.02$ns in the case $\bar n=1$ with
the parameters
$\{\omega_0,\omega,\omega_\text{dr},g,\Omega\}/2\pi=\{10,5,5,0.125,10\}$GHz.
The operating time for the single-qubit rotation along $x$ axis is
$t_x\sim1/\Omega_\text{R}\approx0.3$ns with the above parameters.
The single-qubit rotation along $z$ axis takes a time
$t_z\sim1/\Omega'_\text{R}\approx0.03$ns with the parameters
different from above to obtain the desired evolution of the system,
that is $\omega/2\pi\approx1$MHz (the rest are same). The two-qubit
gate in (\ref{eq:2qHam}) can be realized on the time scaling
$t_{2q}$ which satisfies $g^2t_{2q}/\delta=\pi/4$ and is calculated
as $t_{2q}\approx8$ns with the parameters
$\{g,\delta\}/2\pi=\{0.125,1\}$GHz. Here for the two-qubit gate we
choose the tunneling $T\approx18\mu$eV which was recently realized
in~\cite{Haider09}. Thus, all these operating times are less than
the minimum decoherence time.

Now we analyze the effect on gate operations due to noise. The
variations of the energy gap $\Delta(t)$ caused by the fluctuation
of the electric field would lead to unwanted phase to the desired
gate operations.

We use the two-qubit gate in Eq.~(\ref{eq:2qHam}) as an example.
With the time dependent fluctuations $\delta \lambda(t)$ of the
effective coupling coefficient $\lambda=g^2/\delta$, the evolution
operator of the system becomes $
U'_{2q}=U_{2q}\exp{\big\{-\text{i}\int_{0}^{t_{2q}}\text{d}t\delta\lambda(t)(\frac{3}{2}\sum_{i=1,2}\hat{\sigma}_z^i+\hat{\sigma}_+^1\hat{\sigma}_-^2+\hat{\sigma}_-^1\hat{\sigma}_+^2)\big\}}
$, where the unwanted phase
$\phi=\int_{0}^{t_{2q}}\text{d}t\delta\lambda(t)$. The distribution
of the unwanted phase becomes Gaussian distribution because
$\lambda$ is in Gaussian distribution. With the parameters above, we
numerically calculate the variances of the unwanted phase
$\text{Var}(\phi)\sim5\times10^{-3}\pi$.

For single-qubit $\sigma_x$ gate, the unwanted phase is
$\int_0^{t_x}\text{d}t\delta\Omega_\text{R}(t)$, while for
$\sigma_z$ gate, that becomes $\int_0^{t_z}\text{d}t\delta
\Omega'_\text{R}(t)$. With the same method, we can calculate the
variance of the phases.

From the analysis, we show that even the dephasing occurs over the
gate operation, we can still implement a universal set of gates with
high fidelities. For example, with the parameters we show above the
fidelity for the entangling two-qubit gate is about $0.9946$,
$0.9952$ for $\sigma_x$ gate and $0.9961$ for $\sigma_z$ gate.

\section{Summary}
If a quantum computer is built, intractable
problems such as factorization would be solved efficiently, with
enormous ramifications for communication security. Semiconductor
DDMs quantum computer, which would capitalize on chip fabrication
technology and could be hybridized with existing computers, is the
preferred method for quantum computation. Here we propose scalable
quantum computing with electrically controlled semiconductor spins
of DDMs coupled to a microwave stripline resonator on a chip.
Quantum information is encoded in the singlet states of DDMs.
Initialization of qubits can be realized with an adiabatic passage.
With the switchable coupling to the resonator, we can implement a
universal set of quantum gates on any qubit. Although in general
charge qubits have less coherent life time compared to spin qubits,
the generation and measurement methods are much simpler and faster,
which makes our protocol competitive with spin qubits in the context
of circuit-based quantum computing. Because of the switchable
coupling between the double-dot pairs and the resonator, we can
apply this entangling gate on any two qubits without affecting
others, which is not trivial for implementing scalable quantum
computing and generating large entangled state. The fidelities of
the gates in our protocol are studied including all kinds of major
decoherence, with promising results for reasonably achievable
experimental parameters. The feasibility of this scheme is
characterized through exact numerical simulations that incorporate
various sources of experiment noise and these results demonstrate
the practicality by way of current experimental technologies.

\section{Acknowledgements}
This work has been supported by National Natural Science Foundation
of China, Grant No. 10944005, Southeast University Startup fund,
NSERC, MITACS, CIFAR, QuantumWorks and iCORE.

\end{document}